\documentclass[conference]{IEEEtran}
\IEEEoverridecommandlockouts
\usepackage{amsmath,amssymb,amsfonts}
\usepackage{algorithmic}
\usepackage{graphicx}
\usepackage{textcomp}
\usepackage{array}
\usepackage{arydshln}
\usepackage{bm}
\usepackage[usenames,dvipsnames]{xcolor}
\usepackage{adjustbox}
\usepackage{microtype}
\usepackage{flushend}
\usepackage{mathtools}
\usepackage[sorting=none]{biblatex}
\usepackage{listings}
\usepackage{float}
\usepackage{multicol}
\usepackage[font=footnotesize]{caption}
\usepackage{subcaption}%
\usepackage[justification=centering]{caption}
\usepackage{balance}

\def\BibTeX{{\rm B\kern-.05em{\sc i\kern-.025em b}\kern-.08em
    T\kern-.1667em\lower.7ex\hbox{E}\kern-.125emX}}

\usepackage[pdfusetitle]{hyperref}%
\usepackage[capitalise]{cleveref}%
\usepackage{booktabs}%
\usepackage{csvsimple}%
\usepackage[%
  round-mode=places,%
  round-precision=2,%
  table-format=1.2,%
]{siunitx}%
\usepackage{tikz, pgfplots, pgfplotstable}%
\usepgfplotslibrary{units}
\usetikzlibrary{patterns}
\usepackage[linesnumbered,lined,ruled,commentsnumbered]{algorithm2e}%
\usepackage{listings}
\usepackage{multirow}

\PassOptionsToPackage{hyphens}{url}%

\colorlet{punct}{red!60!black}
\definecolor{background}{HTML}{EEEEEE}
\definecolor{delim}{RGB}{20,105,176}
\colorlet{numb}{magenta!60!black}
\definecolor{accent8c1}{HTML}{7FC97F}

\lstdefinelanguage{json}{
    basicstyle=\scriptsize\ttfamily,
    stepnumber=1,
    numbersep=8pt,
    showstringspaces=false,
    literate=
     *{0}{{{\color{numb}0}}}{1}
      {1}{{{\color{numb}1}}}{1}
      {2}{{{\color{numb}2}}}{1}
      {3}{{{\color{numb}3}}}{1}
      {4}{{{\color{numb}4}}}{1}
      {5}{{{\color{numb}5}}}{1}
      {6}{{{\color{numb}6}}}{1}
      {7}{{{\color{numb}7}}}{1}
      {8}{{{\color{numb}8}}}{1}
      {9}{{{\color{numb}9}}}{1}
      {:}{{{\color{punct}{:}}}}{1}
      {,}{{{\color{punct}{,}}}}{1}
      {\{}{{{\color{delim}{\{}}}}{1}
      {\}}{{{\color{delim}{\}}}}}{1}
      {[}{{{\color{delim}{[}}}}{1}
      {]}{{{\color{delim}{]}}}}{1},
}

\lstdefinelanguage{jsontiny}{
    basicstyle=\tiny,
    frame=single, 
    linewidth=.5\textwidth,
    stepnumber=1,
    numbersep=8pt,
    showstringspaces=false,
    literate=
     *{0}{{{\color{numb}0}}}{1}
      {1}{{{\color{numb}1}}}{1}
      {2}{{{\color{numb}2}}}{1}
      {3}{{{\color{numb}3}}}{1}
      {4}{{{\color{numb}4}}}{1}
      {5}{{{\color{numb}5}}}{1}
      {6}{{{\color{numb}6}}}{1}
      {7}{{{\color{numb}7}}}{1}
      {8}{{{\color{numb}8}}}{1}
      {9}{{{\color{numb}9}}}{1}
      {:}{{{\color{punct}{:}}}}{1}
      {,}{{{\color{punct}{,}}}}{1}
      {\{}{{{\color{delim}{\{}}}}{1}
      {\}}{{{\color{delim}{\}}}}}{1}
      {[}{{{\color{delim}{[}}}}{1}
      {]}{{{\color{delim}{]}}}}{1},
}

\lstdefinelanguage{myPython}{%
  language=Python,%
  classoffset=1,
  morekeywords={parallel, vectorize},%
  classoffset=0,
}

\usetikzlibrary{tikzmark}%
\usetikzlibrary{graphs}%
\usepgfplotslibrary{colorbrewer}%

\pgfplotsset{%
  compat = newest,%
  table/search path = {data},%
  my bargraph colors/.style = {%
    colormap/Paired,%
    cycle list = {%
      {index of colormap={0}, fill=., draw=.},%
      {index of colormap={1}, fill=., draw=.},%
      {index of colormap={2}, fill=., draw=.},%
      {index of colormap={3}, fill=., draw=.},%
      {index of colormap={4}, fill=., draw=.},%
      {index of colormap={5}, fill=., draw=.},%
      {index of colormap={6}, fill=., draw=.},%
      {index of colormap={7}, fill=., draw=.},%
      {index of colormap={8}, fill=., draw=.},%
    },%
  },%
  my bargraph style/.style = {%
    axis on top,%
    axis lines* = left,%
    major grid style = {%
      draw = white,%
    },%
    my bargraph colors,%
    area legend,%
    reverse legend,%
    legend style = {%
      draw=none,%
    },%
  },%
  my xbar style/.style = {%
    xbar = 0pt,%
    bar width = 6pt,%
    xmajorgrids,%
    xtick distance = 1,%
    major x tick style = { draw=none, },%
    major x grid style = { white, },%
    my bargraph style,%
    tickwidth = 0pt,%
    x axis line style = {%
      opacity = 0,%
    },%
    xticklabels = {\empty},%
    y axis line style = {%
      draw = black,%
    },%
    To change tick font sizes:
    every tick label/.append style = {%
      font=\footnotesize{},%
    },%
    ytick = data,%
    y dir = reverse,%
    visualization depends on=rawx\as\rawx,%
    nodes near coords = {%
      \footnotesize{}\pgfmathprintnumber{\rawx}%
    },%
  },%
  polybench speedup style/.style = {%
    width=1.0\textwidth,%
    height=4cm, 
    enlarge x limits = { abs = 1em, },%
    y axis line style = {%
      opacity = 5,%
    },%
    bar width=5pt,%
    ybar=0.5pt,%
    ymajorgrids,
    ytick distance = 50,%
    ymajorticks=true,%
    xtick pos=bottom,%
    legend pos=north east,%
    legend cell align={left},%
    legend style={font=\small},%
    every axis plot/.append style={fill},%
    cycle list/Accent,
  },%
  customOp speedup style/.style = {%
    height=7cm, 
    enlarge x limits = { abs = 1em, },%
    y axis line style = {%
      opacity = 0,%
    },%
    bar width=8pt,%
    ytick pos=left,%
    ybar=0pt,%
    ymajorgrids,
    ytick distance = 1,%
    ymajorticks=true,%
    xtick pos=bottom,%
    xtick=data,%
    xticklabel style = { rotate = 60, anchor = east, font=\small },%
    x label style={at={(axis description cs:0.5,-0.28)},anchor=north},%
    legend pos=north west,%
    legend cell align={left},%
    legend style={font=\small},%
    every axis plot/.append style={fill},%
    cycle list/Accent,
  },%
}%
\pgfplotstableset{%
  col sep=comma,%
}%
\pgfplotsset{compat=1.11,
    /pgfplots/ybar legend/.style={
    /pgfplots/legend image code/.code={%
       \draw[##1,/tikz/.cd,yshift=-0.25em]
        (0cm,0cm) rectangle (3pt,0.8em);},
   },
}

\def\DOM{\mathcal{D}}

\newcommand{\customop}[0]{hybrid custom operators~}

\newcommand{\iter}[0]{\vec{it}}

\begin{document}

\title{PolyTOPS: Reconfigurable and Flexible Polyhedral Scheduler}


\author{
Gianpietro Consolaro\IEEEauthorrefmark{1}\IEEEauthorrefmark{3},
Zhen Zhang\IEEEauthorrefmark{1},
Harenome Razanajato\IEEEauthorrefmark{1},
Nelson Lossing\IEEEauthorrefmark{1},
Nassim Tchoulak\IEEEauthorrefmark{1},\\
Adilla Susungi\IEEEauthorrefmark{1},
Artur Cesar Araujo Alves\IEEEauthorrefmark{1},
Renwei Zhang\IEEEauthorrefmark{2},
Denis Barthou\IEEEauthorrefmark{1},
Corinne Ancourt\IEEEauthorrefmark{3},\\
Cedric Bastoul\IEEEauthorrefmark{1},\\
\IEEEauthorrefmark{1}Huawei Technologies France, Paris, France
\IEEEauthorrefmark{2}Huawei Technologies Co., Ltd., Beijing, China\\
\IEEEauthorrefmark{3}Mines-Paris PSL University 
}

\maketitle
\thispagestyle{empty}
\pagestyle{empty}

\begin{abstract}
Polyhedral techniques have been widely used for automatic code optimization in low-level compilers and higher-level processes.
Loop optimization is central to this technique, and several polyhedral schedulers like Feautrier, Pluto, isl and Tensor Scheduler have been proposed, each of them targeting a different architecture, parallelism model, or application scenario.
The need for scenario-specific optimization is growing due to the heterogeneity of architectures.
One of the most critical cases is represented by NPUs (Neural Processing Units) used for AI, which may require loop optimization with different objectives. 
Another factor to be considered is the framework or compiler in which polyhedral optimization takes place.
Different scenarios, depending on the target architecture, compilation environment, and application domain, may require different kinds of optimization to best exploit the architecture feature set.

We introduce a new configurable polyhedral scheduler, PolyTOPS, that can be adjusted to various scenarios with straightforward, high-level configurations. This scheduler allows the creation of diverse scheduling strategies that can be both scenario-specific (like state-of-the-art schedulers) and kernel-specific, breaking the concept of a one-size-fits-all scheduler approach.
PolyTOPS has been used with isl and CLooG as code generators and has been integrated in MindSpore AKG deep learning compiler. Experimental results in different scenarios show good performance: 
a geomean speedup of 7.66x on MindSpore (for the NPU Ascend architecture) \customop over isl scheduling,
a geomean speedup up to 1.80x on PolyBench on different multicore architectures over Pluto scheduling. Finally, some comparisons with different state-of-the-art tools are presented in the PolyMage scenario.
\end{abstract}

\vspace{1mm}

\section{Introduction}
    The polyhedral model has been widely used in modern optimizing compilers and frameworks for deep learning workloads, e.g., TC (Tensor Comprehension)~\cite{TC} for PyTorch, AKG (Automatic Kernel Generator)~\cite{AKG} for MindSpore~\cite{chen2021deep}, nGraph~\cite{nGraph} and Affine Dialect in MLIR~\cite{MLIR}.
    Loop-based representations of computational kernels, combined with automatic mathematical (affine) transformations, enhance parallelism and data locality efficiency on the target hardware. 
   This technique is characterised by its ability to systematically optimize execution time in most cases without (or with minimal) manual adjustment. It has been demonstrated to be successful on CPU, GPU and NPU (Neural Processing Unit), resulting in impressive performance improvements.
    
    The core of the polyhedral optimization is the polyhedral scheduler, which applies affine transformations on the input for-loops according to its objective function. Some well-known polyhedral schedulers are the \textit{Feautrier} scheduler~\cite{Feautrier}, \textit{Pluto}~\cite{Pluto}, \textit{isl-scheduler}~\cite{isl}\cite{PPCG}, and \textit{Tensor} Scheduler~\cite{TensorScheduler}. They are characterized by the use of different cost functions, optimizing for different specific scenarios.

    Polyhedral scheduler algorithms are based on mathematical optimization.
    Affine constraint systems and cost functions are constructed to maximize hardware efficiency by iteratively solving Integer Linear Programming (ILP) problems to find optimal transformations for each loop dimension. The lack of controllability and configurability makes it challenging to produce efficient transformations for new architectures or scenarios.
    For example, isl was used in the AKG project to target Huawei's NPU \emph{Ascend architecture}~\cite{Ascend}.
    It performs well in many cases but poorly in others.
    For instance, it cannot produce the transformation illustrated in \cref{fig:Optimization_example}, which could easily be lowered on the vector unit of the NPU.
    
\begin{figure}[h]%
  \centering{}%
    \begin{subcaptionblock}{.45\textwidth}%
        \begin{lstlisting}[frame=single, linewidth=.5\textwidth]
for(i = 0; i < 100; i++){
  for(j = 0; j < 10; j++){
0:  c[j][i] = a[j][i] * b; 
1:  d[i][j] = e[i][j] * x; 
  }
}
        \end{lstlisting}%
        \phantomcaption{}%
        \label{lst:optimization-example:original}%
    \end{subcaptionblock}%
    \hfill{}%
    \begin{subcaptionblock}{.45\textwidth}%
        \begin{lstlisting}[frame=single, linewidth=.5\textwidth]
for(j = 0; j < 10; j++)
  for(i = 0; i < 100; i++)
0:  c[j][i] = a[j][i] * b;
    
for(i = 0; i < 100; i++)
  for(j = 0; j < 10; j++)  
1:  d[i][j] = e[i][j] * x; 
        \end{lstlisting}%
        \phantomcaption{}%
        \label{lst:optimization-example:npu}%
    \end{subcaptionblock}%
  \vspace{-2mm}%
  \addtocounter{figure}{-1}%
  \setcaptiontype{lstlisting}%
  \caption{\textbf{Left}: original for-loop code. It is fully parallel and suitable for GPU architecture. The \textbf{Right} one is optimal for vectorization of the NPU (for both vectorized data loading and computing) thanks to the loop distribution and the interchange for the statement \textbf{0}. Pluto or isl may be able to find the loop distribution (specifying the correct fusion heuristic), but no interchange would be found.}%
  \label{fig:Optimization_example}%
\end{figure}

     Improving performance when state-of-the-art schedulers cannot find the optimal transformation can be attempted by adjust the initial scheduling results through additional passes, as can be seen,
     for example, in AutoPoly~\cite{cedrickeynote}, the AKG  module for Ascend backend.
     This method can be cumbersome, considering the complexity of finding optimal transformations that preserve the semantics.
    A recent idea, ``Constraint Injection''~\cite{Mindtrick}, proposed to build an interface for the classical polyhedral scheduler that allows the injection of custom constraints or partial transformations. 
    This approach can drive the scheduler to generate appropriate initial scheduling.
    It shows execution time speed-ups of deep learning workloads on GPU. 
    However, it is kernel-specific and only allows to optimize for the specific input case.
    Generalizing an expected optimization for any input kernel requires more engineering effort on pre-processing, e.g. dependency analysis, pattern matching on memory access, etc.

    This paper proposes a novel design of a fully controllable iterative polyhedral scheduler: \textbf{PolyTOPS}.
    It allows the production of architecture-oriented optimizations (e.g. \cref{fig:Optimization_example}) for any case from simple user configurations, and it can be easily adapted to new scenarios. 
    PolyTOPS innovations are twofold:    
    \begin{itemize}
        \item \textbf{Configurability}: PolyTOPS provides a rich expressivity on schedule strategies specification via an easy-to-use interface.
        All aspects of an iterative scheduling mechanism can be configured, e.g., parallelism control, vectorization, temporal and spatial locality, fine-grained controlling of statements loop fusion and fission, as well as partial schedule specifications.
        Moreover, the behaviour of PolyTOPS can be elegantly changed into a well-defined approach, e.g., Pluto-style, Feautrier-style, isl-style, Tensor-scheduler-style or extended to define novel strategies, e.g., scenario-specific, kernel-specific, extending the idea presented in~\cite{Mindtrick}.

        \item \textbf{Flexibility}:
        Instead of a single ``one-size-fits-all'' method,
        PolyTOPS exhibits a versatile design that can address scenario-specific optimizations.
        It is possible to start from a given generic strategy with little effort and then incrementally adjust this strategy for some particular loops of the kernel or for some particular architecture.
        PolyTOPS provides an extendable infrastructure for an iterative scheduler where constraints can be finely tuned
        -- from predefined strategies down to dedicated transformation heuristics -- for each statement and loop.
        We show how our approach can target multiple architectures (different types of CPU and Ascend NPU) and compare speedups with state-of-the-art schedulers.
    \end{itemize}    
      
    We briefly introduce background notions for polyhedral schedulers in Section~\ref{sec:background}. PolyTOPS design and implementation are detailed in Section~\ref{sec:polytops}, and benchmark results on CPU and NPU are presented in Section~\ref{sec:results}.

\section{BACKGROUND}\label{sec:background}
Polyhedral optimization of kernels can roughly be decomposed into three stages.
First, the input code and loop nests are represented as polyhedra.
Then, algebraic transformations are applied to them,
finally, a new optimized code that scans the transformed polyhedra is generated. In this section, we provide an overview of the techniques used in the first two stages and describe state-of-the-art methods.

\subsection{Polyhedral Model}\label{sec:polyhedralmodel}

\subsubsection{Iteration Domain}
For each statement of the code, the iteration domain represents the range of values taken by the loop iterators surrounding this statement. The vector~$\iter$ composed by these iterators is called the \textit{iteration vector}. Iteration domains are assumed to be polyhedra of iteration vectors and can depend on parameters. The vector of parameters~$\vec{\textit{N}}$ is composed of variables that are constant during the execution of the code.
The domain~$\DOM$ of a statement~$S$ is defined as: 

\begin{equation*}
    \begin{gathered}
      \DOM_{S} =
        \left\{\, 
          \iter
          ~\left|~ 
          M_S \cdot
          \begin{pmatrix*}[c]
            \iter \\
            \vec{N} \\
            1 \\
          \end{pmatrix*} 
          \geq 0
          \right.
        \right\}
    \end{gathered}
\end{equation*}
where \textbf{$M_S$} is a matrix defining the domain polyhedron.

\vspace{0.2cm}
\subsubsection{Dependencies and Legality}
A dependency~$\delta_{S\to R}$ from statement~$S$ to statement~$R$ means that statement~$S$ needs to be executed before statement~$R$ to preserve the semantics of the program. This dependency is defined on a set of iteration vector values for $S$ and $R$, with the following constraints: $S$ and $R$ access to the same memory location (either $S$ or $R$ is a write) and $S$ is executed before $R$. These constraints, similarly to the domain of statements, define a polyhedron:
\begin{equation*}
    \begin{gathered}
      \delta_{S\to R} =
        \left\{\, \left(
          \begin{array}{c}
          \iter_{S}\\
          \iter_{R}
          \end{array}
          \right)
          ~\left|~ 
          M_{S\to R} \cdot
          \begin{pmatrix*}[c]
            \iter_{S} \\
            \iter_{R} \\
            \vec{N} \\
            1 \\
          \end{pmatrix*} 
          \geq 0
          \right.
        \right\}
    \end{gathered}
\end{equation*}

\vspace{0.2cm}
\subsubsection{Scheduling Function}
A scheduling function~$\Theta$ maps each statement and iteration vector of its domain to a unique multi-dimensional date.
Dates are totally ordered with the lexicographic order. 
Given a statement~$S$, $\Theta_S$ is a multidimensional function defined dimension-wise by affine forms $\phi_{S,i}$. These affine forms depend on iterators~$\iter$ and parameters~$\vec{N}$.
$\Theta_S$ can be defined as follows:
   \begin{eqnarray*}
     \begin{gathered}
       \Theta_S :%
         \begin{array}{c}
           \DOM_S( \vec{N})\\
           \iter
         \end{array}%
         \begin{array}{l}
           \rightarrow {\mathbb{N}^m}\\
           \mapsto ( \phi_{S,0}(\iter) ~ ... ~ \phi_{S,m-1}(\iter) )
         \end{array}%
     \end{gathered}%
   \end{eqnarray*} 
where \textit{m} is the number of scheduling dimensions, and $\phi_{S,i}$ are defined by:
   \begin{equation}
   \label{eq:single_schedule_dim}
     \begin{gathered}
       \phi_{S,i}(\iter)  =    T_{S,i} \cdot
                        \begin{pmatrix*}[c]
                        \iter\\
                        \vec{N} \\
                        1 \\
                        \end{pmatrix*} 
     \end{gathered}
   \end{equation}
 where $T_{S,i}$ is the transformation vector.
 
 For a statement~$S$ surrounded by~$k$ nested loops, at most $2k+1$~dimensions~\cite{PouchetMulti} are necessary to express all possible scheduling transformations (strip mining is not expressed through the scheduler), but in practice, there is no upper limit on the number of dimensions.

 \subsection{Scheduler}\label{sec:scheduler}
 The polyhedral scheduler is an algorithm that computes the scheduling function~$\Theta$. Two types of constraints govern the computation of this function. It has to preserve the semantics of the initial code and optimize some cost functions. Both types of constraints are integer affine constraints.  
 
 We now give an overview of the main components necessary to build the ILP problem, with the proximity cost function representing the most used cost function defined in the state-of-the-art:
\vspace{0.2cm}
 \subsubsection{Scheduling Problem Formalization}
 PolyTOPS is an \textit{iterative} scheduler: The algorithm will find the full scheduling transformations~$\Theta$ step by step, building an ILP problem to find each scheduling dimension~$\phi_{S,i}$, starting from the outermost dimension until the innermost dimension. The algorithm makes sure to terminate when enough scheduling dimensions are found.
 The scheduler aims to find the optimal vector of coefficients~$T_{S,i}$ for all~$S$.

\vspace{0.2cm}
\subsubsection{Validity/Legality Constraint} The validity constraint has been introduced by Feautrier~\cite{Feautrier}. This is the core of the polyhedral scheduler because it constrains the transformation vectors~$T_{S,i}$ to have values that preserve the program semantic (ensuring the legality of the schedule). 
For each dependency~$\delta_{S\to R}$, $S$ has to be executed before $R$:
\begin{equation*}
  \begin{gathered}
  (\iter,\iter')\in \delta_{S\to R} \Rightarrow \Theta_{R}(\iter') \succ \Theta_{S}(\iter)
  \end{gathered}
\end{equation*}
where the symbol~$\succ$ stands for lexicographically greater. 

Considering that for an iterative scheduler, each dimension~$\phi_{S,i}$ is computed from the outermost to the innermost, the definition of validity becomes:
\begin{equation}
\label{eq:dep}
  \begin{gathered}
    (\iter,\iter')\in \delta_{S\to R} \Rightarrow \phi_{R,i}(\iter') \geq \phi_{S,i}(\iter)
  \end{gathered}
\end{equation}
until the dependency~$\delta_{S\to R}$ is satisfied. 
This implication can be linearized using the Farkas Lemma~\cite{ilp_theory}\cite{Feautrier}: 
constraints are then expressed only on the space of variables composed by the vectors~$T_{S,i}$ and~$T_{R,i}$.

\vspace{0.2cm}
\subsubsection{Progression Constraint}
The progression constraint is added at each scheduling iteration to ensure the progression of the algorithm.
Its role is to ensure that the schedule defines a complete order for the iteration space and to make sure that the trivial zero solution is avoided.
The constraint definition forces the next scheduling solution to be linearly independent of previous solutions in the iteration space.

We define the matrix~$H_{S}$ as the concatenation (row by row) of previous scheduling dimension solutions~$T_{S,i}$.
We define the orthogonal complement~$H_{S}^{\bot}$ as follows:

\begin{equation*}
   \begin{gathered}
        H_{S}^{\bot} = I - H_{S}^{T}(H_S H_S^{T})^{-1} H_S
   \end{gathered}
\end{equation*}
where~$I$ is the identity and~$H_S^{T}$ is the transposition of~$H_S$.

The progression constraint, considering that we limit our scheduling search space in the positive orthant, is defined as the sum (row by row) of the orthogonal complement matrix as follows, 
\begin{equation}\label{eq:progression}
   \begin{gathered}
        \forall i, H_{S,i}^{\bot}\cdot h_S^* \geq 0 \;\;\;\; \wedge \;\;\;\; \sum_i H_{S,i}^{\bot}\cdot h_S^* \geq 1
    \end{gathered}
\end{equation}
with~$H_{S,i}^{\bot}$ a row of~$H_{S}^{\bot}$, and~$h_S^*$ the next solution to be computed.

\vspace{0.2cm}
\subsubsection{Proximity Cost Function}
\label{Proximity Section}
The Proximity cost function was defined by Bondhugula \textit{et al.}~\cite{Pluto} in order to find among legal solutions the ones that optimize temporal locality.

The idea is to minimize the distance (in scheduling time) between multiple accesses to the same memory position. Data dependences describe multiple accesses to the same memory position, then the \textit{Proximity} objective is to minimize the dependency distance.
For a dependency~$\delta_{S\to R}$, the constraint is defined by:
\begin{equation}
    \begin{gathered}
        (\iter,\iter')\in \delta_{S\to R} \Rightarrow\phi_{i,R}(\iter') - \phi_{i,S}(\iter) \leq \vec{u}\vec{N} + w
    \end{gathered}
\end{equation}
where $\vec{u}$ and $w$ are the cost functions to minimize. 

\textit{Proximity} accurately represents a useful transformation characteristic and, indirectly, favours the first dimensions to be parallel, with a dependency distance of 0.

\subsection{State of the Art}\label{sec:stateoftheart}
In the polyhedral framework defined before, we briefly describe various state-of-the-art schedulers.

    \textit{Feautrier}'s~\cite{Feautrier} scheduler is the first \textit{iterative} polyhedral scheduler. 
    The target is to optimize for single-core SIMD CPUs. 
    The \textit{Validity} constraint is combined with the \textit{Feautrier} cost function. This cost function aims to find sequential outer dimensions that could carry as many dependencies as possible. This can lead to inner loop parallelism for SIMD vectorization exploitation.

    \textit{Pluto}~\cite{Pluto} iterative scheduler introduces the \textit{Proximity} cost function previously described. The aim is to exploit high parallelism in architectures like multi-core CPUs. A more recent version, \textit{Pluto+}~\cite{bondhugula2016pluto-plus}, extends features to support loop reversal and negative skewing and finds the solutions for some corner case problems that could not be solved by \textit{Pluto}. \textit{Pluto-lp-dfp}~\cite{Pluto-lp} is an extension resorting to linear programming instead of ILP. This relaxation decomposes the scheduling algorithm into a sequence of transformations, showing the potential benefits in terms of compilation time.

    \textit{isl}~\cite{isl}\cite{PPCG} iterative scheduler uses both the re-implementations of \textit{Pluto} and \textit{Feautrier} schedulers to maximize parallelism. If no external parallelism is found, the \textit{Feautrier} cost function is applied to remove as many dependencies as possible and to find parallelism in subsequent dimensions.

    \textit{Tensor}~\cite{TensorScheduler} iterative scheduler is applied to tensor-based applications, such as AI, typically characterized by high parallelism and few dependencies. Their focus is the definition of the \textit{Contiguity} cost function for cache spatial locality. It tries to find loop permutations that optimize memory access patterns. It achieves good results but is domain-specific, limiting scheduling transformations to loop interchanges only.

    \textit{One-shot}~\cite{PouchetMulti} is a scheduler that is not iterative and is computed by representing the whole multidimensional transformation~$\Theta(S)$ as a single ILP problem. 
    This formulation of the problem makes it easier to represent global constraints and cost functions over the full schedule~$\Theta$ as opposed to iterative schedulers where constraints and cost functions are usually local to a single scheduling dimension~$\phi_i$.
    However, the large number of variables and constraints leads to scalability issues and extended compilation times. Extensions to the One-shot scheduler~\cite{Kong}\cite{CheliniExt} propose addressing the complexity issue via a dictionary of cost functions and a cache mechanism of previously found optimal solutions.

Current research suggests that existing schedulers are tailored to a specific objective function, targeting some architecture.
Their behaviours are predetermined, and the available options do not provide the \textit{flexibility} needed to achieve good results in different areas, for different architectures, and within different compilers.
In AKG, for instance, the optimization target architectures are diverse, so any of the existing schedulers may be successful in some scenarios but not in others. The state-of-the-art still lacks configurability and flexibility.

Several recent works have explored more configurable approaches:  PolyLingual~\cite{hammer2023polylingual} is a work-in-progress domain-specific language (DSL) for polyhedral schedulers. This DSL offers building block functions and types that ease the design of polyhedral schedulers. Expert knowledge is still required to fully design the scheduler logic, a new scheduler or any of the previously cited schedulers, but the set of scheduling strategies it offers should be very wide. However, the designer has to consider all potential edge cases and guarantee the algorithm's completeness. 

A lower-level approach is to directly define the set of polyhedral transformations for a code, as proposed by Tiramisu~\cite{DLCostModel}. In this case, the scheduler is replaced by an AI-guided search strategy among the combination of loop transformations given by the expert. In Clint~\cite{Clint} instead, a graphic interface allows the application of manual transformations directly to the polyhedron. Chlore~\cite{Chlore/Clay} tackles the explainability problem and, given an input kernel and its transformed version, tries to recover the set of polyhedral transformations necessary to obtain the same transformed version.
Finally
``Constraint Injection''~\cite{Mindtrick} proposes a way to inject simple constraints in the polyhedral scheduler, but it is essentially designed to target kernel-specific optimizations.

\section{PolyTOPS}\label{sec:polytops}
PolyTOPS is a configurable iterative polyhedral scheduler. The objective is to propose a flexible, easy-to-adjust tool to ensure that existing strategies, or a mix between existing strategies (generalizing what isl proposes), can be described with very little effort while still allowing the expert to guide the scheduler more precisely if needed be. To achieve this, PolyTOPS provides a general iterative scheduler scheme, where the strategy can be defined through a configuration file.

\begin{figure}[t]%
  \includegraphics[width=\linewidth]{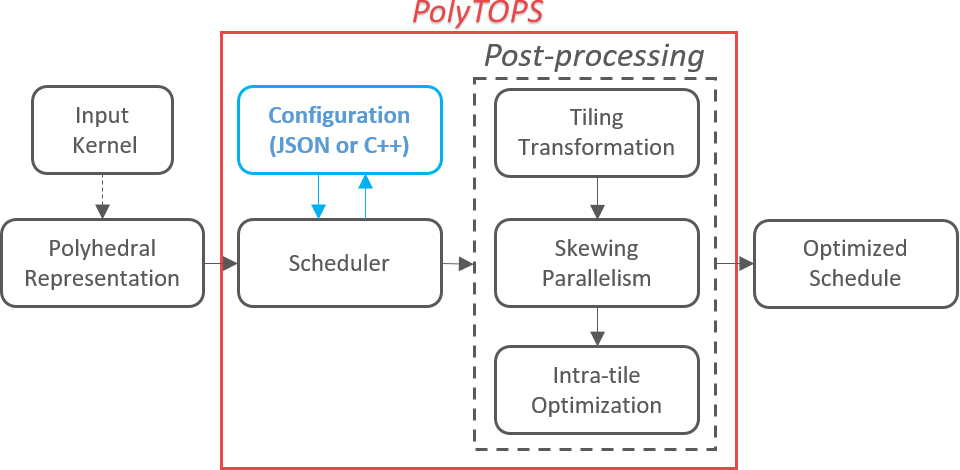}%
  \caption{PolyTOPS workflow representation, showing the major blocks, including the post-processing and the \textbf{configuration}. We support both Openscop and isl-representation as Polyhedral Representation}%
  \label{fig:Polytops}%
\end{figure}

The workflow of PolyTOPS is described in \cref{fig:Polytops}.
The main blocks are similar to schedulers such as isl~\cite{isl} or Pluto~\cite{Pluto}.
Both the input and the output of PolyTOPS are polyhedral representations of the code.
The main parts of the input are the initial schedule and the dependencies.
They can be expressed as isl objects or in OpenScop format.
The result of the core ILP-based scheduling algorithm may be further post-processed:
this phase handles tiling, intra-tile optimization and skewing for wavefront parallelism (see~\cite[Section 5.3]{Pluto}).
It is important to highlight that no tile-size decision is implemented in the core scheduler.
Tile sizes must be externally provided for tiling to be applied.
Finally, as PolyTOPS can output isl objects or an OpenScop representation, the code generation can then be done with tools or libraries such as isl or CLooG~\cite{Cloog}.

The significant contribution of PolyTOPS is in the configuration block, which supports two kinds of interfaces, JSON and C++.
This feature allows the specification of high-level optimization strategies.
Predefined or new strategies can be composed or extended through simple keywords.
Further customization, down to kernel-specific strategies such as statement fusion or partial schedule specification, is also possible.

The configuration of PolyTOPS controls the scheduler for each specific scheduling dimension.
For example, we could add cost functions for a specific dimension, distribute some statements in another dimension or add constraints to another one.
The configurable features can be divided into two main types:
\begin{itemize}
    \item \textit{Local configurations}: They directly control the ILP creation.
    Predefined \textit{cost functions} can be selected, and their priority order can be specified.
    New variables, ILP \textit{constraints} or cost functions can be defined.
    Last but not least, it is possible to define the statement \textit{distribution/fusion} for each dimension. It will be translated internally into specific constraints that will force the distribution specification.

    \item \textit{Global Configurations}: These are higher-level features that do not only impact the definition of the ILP for a specific scheduling dimension.
    These features require several logic steps in the scheduling algorithm to be satisfied.
    For example, the \textit{directives} are suggestions to the scheduler to attempt to vectorize or parallelize a specific loop.
    Another example is \textit{AutoVectorization}, used to automatically detect (based on the memory stride and access pattern) what loops should be scheduled innermost for possible vectorization.
\end{itemize}

Let us now describe all possible configurations, starting from the locals and going on with the global ones.

\subsection{Local Configurations}
\subsubsection{\textit{Cost functions control}}
    A specific combination of predefined or new cost functions and their priorities (given by the textual order, from leftmost to rightmost) can be defined or omitted for each scheduling dimension (\cref{fig:JSONConfiguration} line 7).
    The objective function is the vector of variables.
    The order of the variables is important because they are minimized in lexicographic order. 


New variables can be introduced, as shown in \cref{fig:JSONConfiguration} line 3,
used in \textit{custom constraint} definitions and as cost functions.

The predefined cost functions are \texttt{{proximity}} from Pluto~\cite{Pluto}, \texttt{{feautrier}}~\cite{Feautrier}, \texttt{{contiguity}} inspired from the cost function defined in Tensor scheduler~\cite{TensorScheduler}, and a new simple one named \texttt{{bigLoopsFirst}} that tries to schedule first (outermost) the loops with the largest iteration space.

\begin{figure}[t]%
  \setcaptiontype{lstlisting}%
\begin{lstlisting}[language=json]
1{
2 "scheduling_strategy" : {
3   "new_variables" : ["x"],
4   "ILP_construction" : [
5     {
6       "scheduling_dimension" : "default",
7       "cost_functions" : ["contiguity", "proximity", "x"]
8     }
9   ],
10  "custom_constraints" : [
11    {
12      "scheduling_dimension" : "default",
13      "constraints" : ["x - Si_it_i >= 0"]
14    }
15  ],
16  "fusion" : [
17    {
18      "scheduling_dimension" : 0,
19      "total_distribution" : false, 
20      "stmts_fusion" : [["0", "1"], ["2"]]
21    }
22  ],
23  "directives" : [
24     {
25       "type" : "vectorize",
26       "stmts" : "0",
27       "iterator" : "1"
28     }
29  ]
30 }
31}
\end{lstlisting}%
\caption{\label{fig:JSONConfiguration} JSON example showing most of the configurable features of PolyTOPS, including cost function control and definition, constraints definition, fusion control and directives}%
\end{figure}%

The \texttt{contiguity} cost function is designed to schedule the iterators in the order that offers a better spatial locality.
For each statement~$S$, given the set of iterators~$\iter_S$, we define a cost function named \texttt{contiguity} as follows:
    \begin{eqnarray}
    \label{eq:Contiguity}
            \texttt{contiguity}(S) &=& \sum_{i=0}^{|T_S^{it}|} T_{S,i}^{\vec{it}} \times c_{S,i}\\
            \texttt{contiguity}(S) &\geq &0 \nonumber
    \end{eqnarray}
where~$c_{S,i}$ is a support coefficient that describes a priority order optimizing the memory access pattern.
For instance, focusing on the kernel in \cref{fig:Optimization_example}(left), the two vectors of support coefficient~$\vec{c}_{S0}$ and~$\vec{c}_{S1}$ would be respectively:
    \begin{equation*}
      \begin{gathered}
      \begin{tabular}{c}
            $\vec{c}_{S0} = ( 10 ~~1)$  \\
            $\vec{c}_{S1} = ( 1 ~~10)$
      \end{tabular}
            \left.
            \text{with} ~~\iter_{S0} =  \iter_{S1} =
            \begin{pmatrix*}[c]
                        i\\
                        j \\
            \end{pmatrix*} \notag
            \right.
      \end{gathered}
    \end{equation*}
to force the scheduler to select the outermost loops with the smallest contiguity coefficients. 

\vspace{2mm}
The \texttt{bigLoopsFirst} (BLF) cost function is designed to schedule the loops with the largest domains outermost. The design is the same as the \texttt{contiguity} cost function, but the coefficients $c_i$ used in Formula~\ref{eq:Contiguity} are based on prioritizing the dimensions with the highest bounds first. BLF can be useful in scenarios where kernels have a lot of parallelism, in which only one level or a few levels of outer parallelism are exploitable by the architecture. We try then to maximize the number of parallel iterations.
For instance, focusing on the kernel in \cref{fig:Optimization_example}, the two vectors~$\vec{c}_{S0}$ and~$\vec{c}_{S1}$ would be respectively:

    \begin{equation}
      \begin{gathered}
      \begin{tabular}{c}
            $\vec{c}_{S0} = ( 1 ~~10)$  \\
            $\vec{c}_{S1} = ( 1 ~~10)$
      \end{tabular}
            \left.
            \text{with} ~~\iter_{S0} =  \iter_{S1} =
            \begin{pmatrix*}[c]
                        i\\
                        j \\
            \end{pmatrix*} \notag
            \right.
      \end{gathered}.
    \end{equation}

\vspace{0.2cm}
\subsubsection{\textit{Custom Constraints}}
Using a simple interface, custom constraints are affine inequalities or equations. They can constraint the scheduling functions~$\phi_{S,i}(\iter)$ defined by \cref{eq:single_schedule_dim}, for any statement~$S$ and dimension~$i$ through their vector of coefficients, $T_{S,i}$. We can separate this vector into subvectors $(T_{S,i}^{\iter}\; T_{S,i}^{\vec{N}}\; T_{S,i}^1)$. A constraint can involve any of the coefficient of $T_{S,i}$ using the notation:
\begin{equation*}
    \begin{gathered}
        S[stmt]\_[var\_type]\_[idx\_var],
    \end{gathered}
\end{equation*}
where $i$ is implicitly defined as the current dimension considered (iterative scheduler) and:
\begin{itemize}
    \item \textit{stmt} is a statement number from $0$ to $M-1$ (where $M$ is the number of statements), following the initial textual order. It identifies a unique statement $S$. 
    \item \textit{var\_type} can be one of the following keywords: \textit{it} refers to the subvector $T_{S,i}^{\iter}$, \textit{par} refers to the subvector $T_{S,i}^{\vec{N}}$ and \textit{cst} refers to the constant term $T_{S,i}^1$.
    \item \textit{idx\_var} is the index of the variable. The outermost iterator of a statement is considered iterator 0. The order of the parameters is the textual order in the input program.
\end{itemize}
Additionally, any user-defined variable can be used in the constraints. 
Notice that replacing \textit{stmt} or \textit{idx\_var} with the keyword `\texttt{i}' represents the sum of all the variables of that type.

For example, a constraint that disables skewing for Statement $3$ would be  expressed by:
\begin{equation*}
    \begin{gathered}
        S3\_it\_i \leq 1.
    \end{gathered}
\end{equation*}
This is equivalent to:
 \begin{equation*}
     \begin{gathered}
         \sum_k T_{S3,k}^{\iter} \leq 1.
     \end{gathered}
 \end{equation*}
Custom constraints can be defined either for specific scheduling dimensions or for all of them using the \texttt{fusion} keyword (see the example in \cref{fig:JSONConfiguration} lines 10-15).

 The constraints that are accepted must be \textit{affine}. This means that given the vector of variables~$\vec{V}$ just described, it is possible to define all the constraints in the form:
    \begin{equation*}
      \label{eq:Custom Constraints rules}
        \begin{gathered}
          constraint = ~~ A\vec{V} + c
            \left.
              \begin{tabular}{c}
                $\geq$\\
                $=$\\
              \end{tabular}
            \right.
            0
        \end{gathered}
    \end{equation*}
 where A is a matrix of integer coefficients, and c is an integer.

\vspace{0.2cm}
\subsubsection{\textit{Fusion/Distribution control}}
Custom loop-fusion decisions can be specified for a specific kernel. We give the ability to control the fusion, selecting which statements to fuse and which ones to distribute for each level. \Cref{fig:JSONConfiguration}, lines 16-22, shows a configuration example specifying that statements 0 and 1 are to be fused and statement 2 distributed at the scheduling dimension 0.

\subsection{Global Configurations}

\begin{figure}[t]%
  \setcaptiontype{lstlisting}%
  \centering{}%
\begin{lstlisting}[breaklines=true]
StrategyInfo strategy(last_solution, stmts, old_strategy){
    StrategyInfo new_strategy;
    if(!last_solution.parallel &&
       !old_strategy.recompute_last_solution) {
      new_strategy.recompute_last_solution = true;
      new_strategy.cost_functions = {"feautrier"}
    } else {
      new_strategy.cost_functions = {"proximity"};
    }
    return new_strategy;
}
\end{lstlisting}%
  \caption{\label{fig:ISLConfiguration} An isl style configuration described using the C++ configuration. Feautrier is used as the fallback case when the Proximity fails to extract parallelism}%
\end{figure}%

\subsubsection{\textit{Directives}}
Directives, \cref{fig:JSONConfiguration} lines 23-29, specify that certain loops should be 
\texttt{parallel}, or \texttt{vectorized} (scheduled innermost and not fused) or \texttt{sequential}.
This can be used to suggest partial code transformations while the remaining scheduling transformation decisions are left to the scheduler.
The scheduler will try to satisfy the directives unless scheduling legality can not be guaranteed.
Directives that prevent legality preservation are discarded.

\vspace{0.2cm}

\subsubsection{\textit{Auto Vectorization}} 
This option instructs the scheduler to use a simple heuristic to detect dimensions that could be vectorized for each statement.
The heuristic looks for dimensions that move contiguously in memory.
The scheduler then computes a scheduling transformation where: (1) the vectorizable dimensions are scheduled as innermost and (2) the corresponding statements are unfused for this scheduling dimension.
For architectures such as CPUs or NPUs, vectorization is critical for performance.

\subsection{Configurations Strategy}
The configurations can be specified using two different interfaces, each of them more suitable to different configuration scenarios:
\vspace{0.2cm}
\subsubsection{\textit{JSON interface}}
     The JSON interface, as seen in \cref{fig:JSONConfiguration}, allows to tailor strategies for the input kernel.
     Local configurations are statically defined and mapped to scheduling dimensions.
     The configurations specify cost functions, extra constraints and possible loop distributions.
     However, this interface does not offer the freedom to define complex strategies that take outermost partial schedules into account.

\vspace{0.2cm} 
\subsubsection{\textit{C++ interface}}
     In this configuration, the strategy is defined in a dynamic library that is loaded by PolyTOPS and called before each scheduling iteration.
     This enables a dynamic specification of each scheduling strategy, generalizing isl~\cite{isl} strategy, which calls a Pluto-style scheduler as default and a Feautrier-style scheduler as fallback.
     This example is shown in \cref{fig:ISLConfiguration}.
     Furthermore, the strategy definition has access to many details concerning the statements and the partial schedule computed until the present iteration.
     This gives the opportunity to create more complex strategies.

 The configuration is expressive enough to allow switching between different strategies like Pluto-style, Feautrier-style, isl-style, and TensorScheduler-style and define new ones. The only limit is that the configuration can only influence the core ``Scheduler'' block of \cref{fig:Polytops}.
 For instance, the main Pluto ILP strategy can be easily replicated using the configuration, but the post-processing and internal fusion heuristics cannot.

\begin{algorithm}[t]
\small
\BlankLine
  \KwData{Input Dependencies~$deps$, Statements~$S$, Scheduling Configuration~$\mathit{config}$}
  \KwResult{Scheduling $\Theta$ $\forall S$, Tilability: \textbf{Bands},
  Parallelism info for each level: \textbf{ParallelDimension}}
  \BlankLine

  $\mathit{constraints} \gets \mathit{CreateConstraints(config, deps)}$\;
  $\mathit{dimension} \gets 0$\;
  $band \gets 0$\;
  \Repeat{$P = \emptyset$ \&\& $deps = \emptyset$} {
    \If{ $\mathit{config.type}$ = C++ }{ 
        $\mathit{config} \gets \mathit{UpdateConfiguration(\Theta)}$\;
    }
    \eIf{$\mathit{config.Distribute(dimension)}$}{
        $\phi \gets \mathit{Distribute(dimension, config)}$\;
        $\Theta.Append(\phi)$\;
        $Bands.Append(band)$\; 
        $\mathit{RemoveSatisfiedDependencies(deps)}$\;
        \tcc{Ends the current band}
        $band \gets band+1$\;
    }{
        $ILP \gets \mathit{constraints(dimension)}$\;
        $\phi \gets Solve(ILP)$\;
        \eIf{ $\phi \neq \emptyset$ }{
            $\Theta.Append(\phi)$\;
            \tcc{Same band as before}
            $Bands.Append(band)$\;
        }{
            \tcc{Change band and retry!} 
            $\mathit{RemoveSatisfiedDependencies(deps)}$\;
            $\mathit{band} \gets band+1$\; 
            $\mathit{ILP} \gets \mathit{constraints(dimension)}$\;
            $\phi \gets \mathit{Solve(ILP)}$\;
            \eIf{ $\phi \neq \emptyset$ }{
               $\Theta.Append(\phi)$\;
               $Bands.Append(band)$\;
            }{
                $\phi \gets \mathit{UnfuseSCCs(deps)}$\;
                $\Theta.Append(\phi)$\;
                $\mathit{Bands.Append(band)}$\;
                $\mathit{RemoveSatisfiedDependencies(deps)}$\;
                $band \gets band+1$\;
            }
        }
    }
    $\mathit{ParallelDimension.Append}(\phi.isParallel())$\;
    $P \gets \mathit{ProgressionConstraint}(\Theta)$\;
  }
  return $\Theta$\;
  \caption{PolyTOPS Scheduler}
  \label{alg:scheduling}
\end{algorithm}

\subsection{Common Algorithmic Structure}
PolyTOPS relies on an algorithmic structure shown in Algorithm~\ref{alg:scheduling} that is common to the iterative schedulers, such as Feautrier's~\cite{Feautrier}, Pluto~\cite{Pluto}, isl-scheduler~\cite{isl} and Tensor Scheduler~\cite{TensorScheduler}). This is a generalization of Pluto algorithm, using the configuration strategy to drive the scheduler. 

The termination criteria of the algorithm are to check if the iteration space is completely covered and if all the dependencies are fulfilled (\textit{line 42}). The algorithm iterates to find a new scheduling dimension until the termination criteria are met (from \textit{line 4 to line 42}). To compute the next dimension~$\phi$, the scheduler firstly verifies if the fusion heuristic (or the interface for PolyTOPS) imposes a loop distribution for this scheduling dimension (\textit{lines 8-14}). If not, the algorithm continues with the standard step (\textit{lines 16-21}), constructing the ILP system composed of the cost functions and constraints defined for the dependencies that are not yet completely satisfied. If no solution is found, the algorithm attempts to remove the dependencies satisfied by the previous scheduling dimension, and it continues building the ILP problem and trying to find a solution (\textit{lines 23-30}).
If all preceding steps fail, loop distribution is enforced by analyzing the strongly connected components (SCC) of the dependency graph and distributing the loop of different SCC (\textit{lines 32-36}).
Once the solution is found, it updates the progression constraint, ensuring that the next computed dimension of~$\phi$ will be linearly independent from the previous ones and that the schedule is a bijective transformation.
The algorithm computes $Bands$ and $ParallelDimension$. $Bands$ are used in post-processing tiling to determine which dimensions can be tiled. $ParallelDimension$ indicate which scheduling dimensions are parallel.

This algorithmic scheme covers all iterative schedulers of the literature just by defining the appropriate configurations. PolyTOPS extends them with the ability to select and define the cost functions and constraints (\textit{lines 16, 26}). The JSON interface is expressed statically, so it is parsed once at the beginning of the scheduling algorithm, while the \textit{C++} interface allows for a logic-based decision using the information from the schedule $\Theta$ found so far (\textit{line 6}), so it is updated for each scheduling iteration. For both cases, the configuration impacts loop fusion/distribution decision (\textit{line 9}). 

Legality constraints (\cref{eq:dep}) and progression constraints (\cref{eq:progression}) are always included when computing a solution, whatever the configuration provided. This implies that the scheduler always terminates (similar proof as the one for Pluto \cite{Pluto}). Moreover, the scheduler is guaranteed to find a valid schedule if no custom constraints and no fusion/distribution control are defined in the configuration. Indeed, strategies do not prevent finding a legal schedule, and directives are ignored when they conflict with the legality. Only the custom constraints and fusion/distribution may lead to an empty solution. This is different from approaches such as Tiramisu~\cite{DLCostModel} or other approaches that do not use a scheduler since, in this case, each scheduling function obtained by composing transformations has to be proved valid.

\section{Experimental Results}\label{sec:results}
Our experiments focus on demonstrating the \textit{flexibility} of PolyTOPS, capable of adapting to the different scenarios shown in the following session, and the expressiveness of the \textit{configurability}, capable of changing the behaviour of the scheduler.

\begin{table}[t]
  \vspace{1mm}
  \caption{\textbf{(Ascend 910 NPU)} Custom Operator results, showing the number of cycles for each case and the speedup obtained by PolyTOPS results over the isl ones}
  \centering{}%
  \begin{tabular}{lcrrr}%
    \toprule{}%
      Case &
      Input/Output &
      \begin{tabular}{@{}c@{}}isl \\ (cycles)\end{tabular} &
      \begin{tabular}{@{}c@{}}PolyTOPS \\ (cycles)\end{tabular} &
      Speedup \\
    \midrule{}%
      \begin{tabular}{@{}l@{}}LU \\ decomp\end{tabular}
      & 16x16 & 27943 & 18333 & 1.52 \\
    \midrule{}%
    \multirow{7}{*}{\begin{tabular}{@{}l@{}}trsmL \\ off diag\end{tabular}}
      & 16x16x16  &  15375 &  704 & 21.84 \\
      & 16x16x32  &  31126 & 1122 & 27.74 \\
      & 16x16x48  &  45172 & 1518 & 29.76 \\
      & 16x16x64  &  62414 & 1938 & 32.21 \\
      & 16x16x80  &  75611 & 2324 & 32.53 \\
      & 16x16x96  &  93387 & 2724 & 34.28 \\
      & 16x16x112 & 108384 & 3223 & 33.63 \\
    \midrule{}%
    \multirow{7}{*}{\begin{tabular}{@{}l@{}}trsmU \\ transpose\end{tabular}}
      & 16x16x16  &  55370 &  22100 & 2.51 \\
      & 16x32x16 & 107159 &  44298 & 2.42 \\
      & 16x48x16  & 160547 &  64281 & 2.50 \\
      & 16x64x16  & 212907 &  87914 & 2.42 \\
      & 16x80x16  & 267627 & 106479 & 2.51 \\
      & 16x96x16  & 317589 & 130221 & 2.44 \\
      & 16x112x16 & 370941 & 151204 & 2.45 \\
    \bottomrule{}%
  \end{tabular}%
  \vspace{-3mm}
  \label{CustomOPResults}
\end{table}

\subsection{MindSpore Hybrid Custom Operators}
In the first scenario, PolyTOPS is used in the context of MindSpore~\cite{chen2021deep} for the generation of custom operators on NPU  \customop~\cite{mscustomop} for AI applications. The experiments are run on an Atlas 800 (model 9010) server featuring $8$ Ascend 910 NPU accelerators~\cite{LIANG202075}\cite{liao2021ascend}. The goal is to define custom operators differently from the default AI operators that are already predefined. When creating these operators, it is possible to express some \textit{directives} passed through the AKG~\cite{AKG} compiler to PolyTOPS as part of the internal configuration. PolyTOPS schedules the operator and tries to comply with the provided directives. \cref{CustomOPResults} shows the speedups obtained with such directives compared to isl (the default scheduler previously used in AKG~\cite{AKG}), both cases implemented in MindSpore and isl is used for code generation.  The speedups are significant for all the 3 operators with all the different sizes, with a geomean speedup of 7.66x.

\begin{figure}[t]%
  \small{}%
  \centering{}%
  \begin{subfigure}{\columnwidth}%
\begin{lstlisting}[breaklines=false, language=myPython]
def trsmL_off_diag(a, b):
  inverse_0 = allocate(b.shape, b.dtype)
  row = b.shape[0]
  col = b.shape[1]
  for i in range(row):
    for j in range(i):
      for l in parallel(col // 16):
        for k in vectorize(16):
          inverse_0[i, l*16+k] = a[i, j] * b[j, l*16+k]
          b[i, l*16+k] = b[i, l*16+k] - inverse_0[i, l*16+k]
  return b
\end{lstlisting}%
    \vspace{-2ex}%
    \caption{Input code. Directives are displayed in red.}%
    \label{fig:customop:input}%
  \end{subfigure}%
  
  \begin{subfigure}{\columnwidth}%
\begin{lstlisting}[breaklines=false, language=myPython]
def trsmL_off_diag(a, b):
  inverse_0 = allocate(b.shape, b.dtype)
  row = b.shape[0]
  col = b.shape[1]
  for l in parallel(col // 16):
    for i in range(row):
      for j in range(i):
        for k in vectorize(16):
          inverse_0[i, l*16+k] = a[i, j] * b[j, l*16+k]
        for k in vectorize(16):
          b[i, l*16+k] = b[i, l*16+k] - inverse_0[i, l*16+k]
  return b
\end{lstlisting}%
    \vspace{-2ex}%
    \caption{Optimized code (before tiling) thanks to PolyTOPS}%
    \label{fig:customop:optimized}%
  \end{subfigure}%
  \addtocounter{figure}{-1}%
  \setcaptiontype{lstlisting}%
  \caption{\label{fig:CustomOP}Custom Operator example.}%
\end{figure}

These results come from a manual specification, mostly focusing on vectorization directives that will end up applying interchanges and vectorizing innermost.
We can see an example of one of the operators in \cref{fig:customop:input}.
In this case, the directives hint to vectorize the loop $k$.
The result obtained is shown in \cref{fig:customop:optimized}.
The speedup obtained is because isl would detect $k$ as parallel and schedules it as the outermost loop, thus losing the vectorization opportunity.

Although these results are achieved through manual directive specifications, we discovered that the same configuration file, enabling auto-vectorization and using the proximity cost function, could systematically be used for all kernels for all sizes. This suggests that the ability, through configurations, has the potential not only to obtain kernel-specific optimizations but also to generate effective heuristics for groups of use cases or specific scenarios.

\subsection{Comparing scheduling strategies on Polybench}
The second part of our experiments is focused on the Polybench~\cite{pouchet2012polybench} benchmark. In this experimental section, we chose to compare PolyTOPS results against Pluto. This section uses CLooG~\cite{Cloog} for code generation for all schedulers.

We repeated the tests in three different system configurations: 
\begin{itemize}\label{sec:results:configuration}
    \item \textbf{AMD}: AMD EPYC 7452, with 32 cores (2 threads for each core), 2 sockets. 256 MiB of L3 cache. The compiler version is gcc-11.3.
    \item \textbf{Intel1}: Intel Xeon E5-2683 CPU (x86\_64), with 2 sockets with 16 cores each (2 threads for each core). 80 MiB of L3 cache. The compiler is gcc-10.5.
    \item \textbf{Intel2}: Intel Xeon Silver 4215 CPU (x86\_64), with 2 sockets with 8 cores each (2 threads for each core). 22 MiB of L3 cache. The compiler is gcc-10.5.
\end{itemize}

Polybench contains heterogeneous kernels coming from different domains, such as linear algebra, data mining and stencil computation, and it represents a reference for the polyhedral optimization benchmarks.
In our experiments, the performance obtained by PolyTOPS (using different configurations) is compared to Pluto, using the last development version (commit eddc385). For Pluto, the options \texttt{--parallel} \texttt{--tile} \texttt{--nounrolljam} \texttt{--no-diamond-tiling} are used. The last two are disabled because their post-processing is not available in PolyTOPS so far.

Our study of PolyTOPS showcases three general strategies for each kernel: The proximity cost function is used in the \textit{pluto-style} strategy (\cref{fig:JSON_pluto_tensor} left), while contiguity cost function is used in the \textit{tensor-scheduler-style} strategy (with proximity as secondary). The no-skewing constraint is also applied (\cref{fig:JSON_pluto_tensor} right). The isl-style strategy (\cref{fig:ISLConfiguration}) defaults to proximity, and if no parallelism is found, it recomputes the scheduling dimension, resorting to Feautrier's strategy. These strategies are the same as their state-of-the-art counterparts regarding ILP construction and the primary objective. Our heuristic for the fusion strategy distributes statements with a different loop dimensionality (number of surrounding loops), similar to Pluto's smartfuse heuristic.
\begin{figure}[t]
  \setcaptiontype{lstlisting}%
\begin{minipage}{.45\textwidth}
\begin{lstlisting}[language=jsontiny]
{
 "scheduling_strategy" : {
  "ILP_construction" : [
   {
    "scheduling_dimension":"default",
    "cost_functions":[proximity"]
   }
  ],
 }
}
\end{lstlisting}
    \end{minipage}\hfill
    \begin{minipage}{.45\textwidth}
\begin{lstlisting}[language=jsontiny]
{
 "scheduling_strategy" : {
  "ILP_construction" : [
   {
    "scheduling_dimension":"default",
    "cost_functions":["contiguity",
                      "proximity"],
    "constraints":["no-skewing"]
   }
  ],
 }
}
\end{lstlisting}
    \end{minipage}
\vspace{-2mm}
\caption{\label{fig:JSON_pluto_tensor} JSON configurations showing pluto-style (on the left) and tensor-scheduler-style (on the right)}
\end{figure}

\begin{figure*}[t]
  \centering{}%
    \begin{tikzpicture}%
      \begin{axis}[%
        polybench speedup style,%
        height=4.2cm, 
        ymin = 0.031,%
        ymax = 16,%
        ytick distance = 2,%
        ylabel=Speedup AMD,%
        ylabel style = { font=\footnotesize },
        bar width=3pt,
        legend columns=-1,
        xtick=data,%
        xticklabels=\empty,
        ymode=log,%
        log basis y={2}%
      ]%

        \draw [help lines, dashed,ystep=20, xstep=1,  xshift=3.3mm] (0,0.03) grid(30,18);
        \addplot table [x expr=\coordindex, y=pluto-like, col sep=comma]{Results/Result_Revisioned_Paper/AMD.csv};%
        \addplot table [x expr=\coordindex, y=tensor-scheduler-like, col sep=comma]{Results/Result_Revisioned_Paper/AMD.csv};%
        \addplot table [x expr=\coordindex, y=isl-like, col sep=comma]{Results/Result_Revisioned_Paper/AMD.csv};%
        \addplot[color=blue!50] table [x expr=\coordindex, y=best, col sep=comma]{Results/Result_Revisioned_Paper/AMD.csv};%

        \addlegendentry{pluto-style};%
        \addlegendentry{tensor-scheduler-style};%
        \addlegendentry{isl-style};%
        \addlegendentry{kernel-spec};%
      \end{axis}%
    \end{tikzpicture}%
\vspace{-5mm}
    \begin{tikzpicture}%
      \begin{axis}[%
        polybench speedup style,%
        height=4.2cm, 
        ymin = 0.062,%
        ymax = 16,%
        ytick distance = 2,%
        ylabel=Speedup Intel1,%
        ylabel style = { font=\footnotesize },
        bar width=3pt,
        legend columns=-1,
        xtick=data,%
        xticklabels=\empty,
        ymode=log,%
        log basis y={2}%
      ]%

        \draw [help lines, dashed,ystep=20, xstep=1,  xshift=3.3mm] (0,0.03) grid(30,18);
        \addplot table [x expr=\coordindex, y=pluto-like, col sep=comma]{Results/Result_Revisioned_Paper/intel1.csv};%
        \addplot table [x expr=\coordindex, y=tensor-scheduler-like, col sep=comma]{Results/Result_Revisioned_Paper/intel1.csv};%
        \addplot table [x expr=\coordindex, y=isl-like, col sep=comma]{Results/Result_Revisioned_Paper/intel1.csv};%
        \addplot[color=blue!50] table [x expr=\coordindex, y=best, col sep=comma]{Results/Result_Revisioned_Paper/intel1.csv};%

        \addlegendentry{pluto-style};%
        \addlegendentry{tensor-scheduler-style};%
        \addlegendentry{isl-style};%
        \addlegendentry{kernel-spec};%
      \end{axis}%
    \end{tikzpicture}%
\vspace{-5mm}
    \begin{tikzpicture}%
      \begin{axis}[%
        polybench speedup style,%
        height=4.2cm, 
        ymin = 0.054,%
        ymax = 16,%
        ytick distance = 2,%
        ylabel=Speedup Intel2,%
        ylabel style = { font=\footnotesize },
        bar width=3pt,
        legend columns=-1,
        xticklabels from table={Results/Result_Revisioned_Paper/intel2.csv}{name},%
        xtick=data,%
        xticklabel style = { rotate = 60, anchor = east, font=\footnotesize },%
        ymode=log,%
        log basis y={2}%
      ]%
      
        \draw [help lines, dashed,ystep=100, xstep=1,  xshift=3.3mm] (0,0.006) grid(30,18);
        \addplot table [x expr=\coordindex, y=pluto-like, col sep=comma]{Results/Result_Revisioned_Paper/intel2.csv};%
        \addplot table [x expr=\coordindex, y=tensor-scheduler-like, col sep=comma]{Results/Result_Revisioned_Paper/intel2.csv};%
        \addplot table [x expr=\coordindex, y=isl-like, col sep=comma]{Results/Result_Revisioned_Paper/intel2.csv};%
        \addplot[color=blue!50] table [x expr=\coordindex, y=best, col sep=comma]{Results/Result_Revisioned_Paper/intel2.csv};%

        \addlegendentry{pluto-style};%
        \addlegendentry{tensor-scheduler-style};%
        \addlegendentry{isl-style};%
        \addlegendentry{kernel-spec};
      \end{axis}%
    \end{tikzpicture}%
\vspace{-3mm}
\caption{Speedups (in log scale) of PolyTOPS (using 4 different configurations, \textit{pluto-style}, \textit{tensor-scheduler-style}, \textit{isl-style} and \textit{kernel-specific}) compared to Pluto. The \textit{kernel-specific} configuration is at least as good as the three previous ones. The results are sorted by decreasing \textit{kernel-specific} speedups in Intel2 machine. Tests done on AMD (top), Intel1(middle) and Intel2(bottom)}
\label{fig:BestPolybench}
\end{figure*}
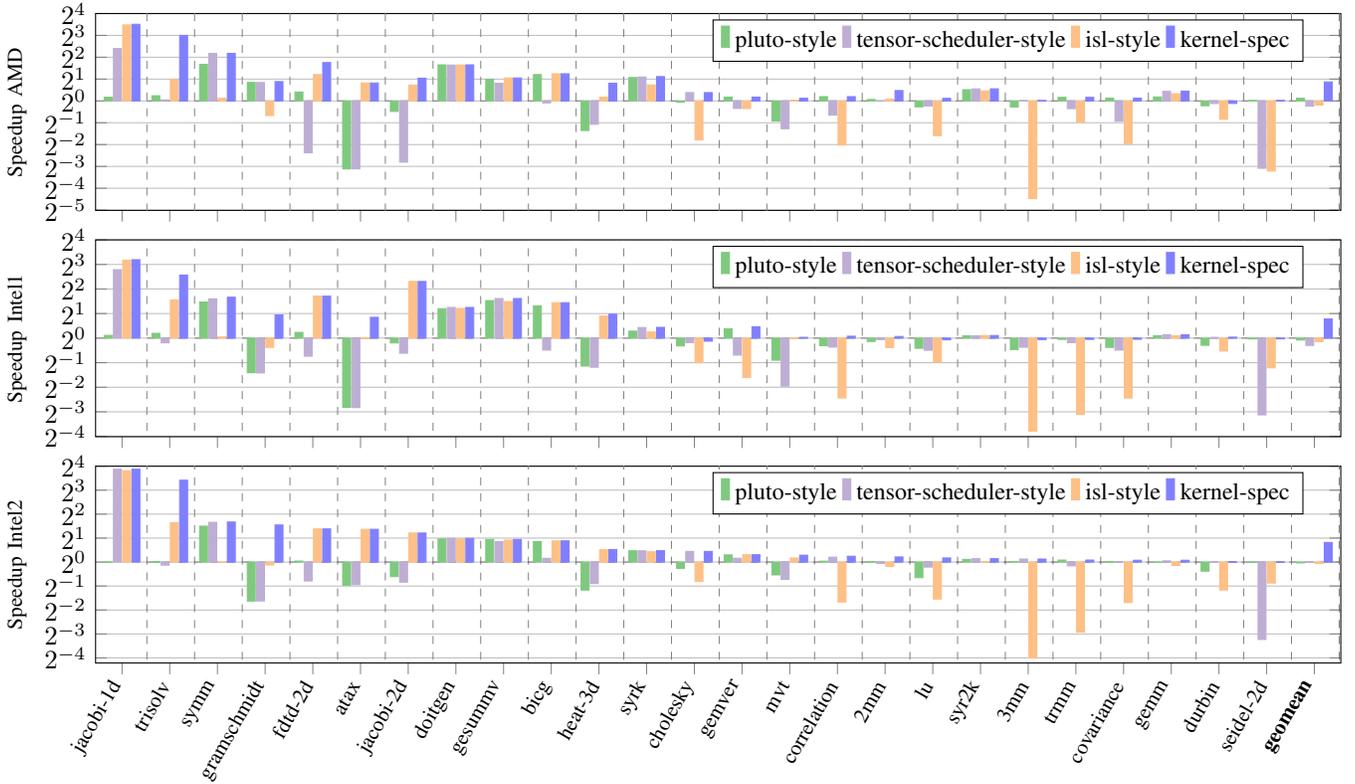

Last but not least, we show the result obtained using a \textit{kernel-specific} configuration for each kernel. These configurations are obtained by playing with the cost functions, fusion decisions and vectorization directives, and they can change between different architectures and kernels.

Out of clarity, in \cref{fig:BestPolybench}, we removed the kernels nussinov, adi, deriche, ludcmp and floyd-warshall where the results are identical between Pluto and PolyTOPS. For the first 4 cases, both Pluto and PolyTOPS fall back to the initial schedule. Performance can be improved 
but it requires support for the negative scheduling coefficients.  Floyd-warshall is too simple to obtain speedups applying loop transformations.

\subsubsection{Results Analysis}
Focusing on the charts in \cref{fig:BestPolybench}, we can see those heuristics like pluto-style, tensor-scheduler-style, and isl-style perform differently depending on the kernel. For example, isl-style performs well for stencil applications with complex dependencies like jacobi-2d, jacobi-1d, and heat-3d, where the Feautrier's fallback is crucial for parallelism. However, in other cases like correlation, covariance, durbin, lu, and trmm, isl-style performs poorly. Pluto-style and tensor-scheduler-style differ mainly in the no-skewing constraint. Pluto-style finds a complex skewing that enables parallelism in jacobi-1d, but the generated code is complex, degrading the overall performance compared to the tensor-scheduler-style solution. On the other hand, pluto-style outperforms tensor-scheduler-style in fdtd-2d because parallelism (that requires skewing) is crucial for performance improvements in this case. In some cases, tensor-scheduler-style performs better because of contiguity interchange.

As expected, the kernel-specific configuration outperforms or at least obtains the same speedup as the other strategies, obtaining an overall \textit{geomean speedup} of \textbf{1.82} for \textit{AMD}, \textbf{1.71} for \textit{Intel1} and \textbf{1.76} for \textit{Intel2}.

For \textit{gramschmidt} (Intel1 and Intel2), in the kernel-specific configuration, thanks to a fusion decision based on maximizing the data reuse, we can find a better speedup compared to Pluto. A hardware counter analysis shows indeed a smaller number of L3-cache-misses (around 5 times less) for our configuration compared to Pluto's.

Another case where fusion is important is showcased in \textit{symm}: Our fusion heuristic decides to distribute one statement from the beginning, enabling parallelism. The result produced by Pluto is, instead, fully sequential (a complete fusion is applied).

Another factor to highlight is the fact that, for a few cases, we need to change the kernel-specific configuration between different architectures. This can be explained by several factors, such as different cache sizes, different numbers of cores and threads and different environments (compiler, architecture, operating system). Among these cases, we can find \textit{jacobi-2d}, \textit{heat-3d} and \textit{fdtd-2d}, where for the Intel machines isl-style is the most performant configuration, while for AMD a simple loop distribution performs better.

In some cases, our pluto-style strategy can outperform the Pluto scheduler, and in some other cases, the reverse situation applies. This is mostly given by different fusion heuristics implemented in the two schedulers. This gives an idea of the impact of the fusion heuristic on the optimization problem and the limits of the existing heuristics.

 These experiments show that kernel-specific configurations can be really useful to explore transformations with minimal effort. However, the results also show that the generic strategies defined so far have room for improvement because they ignore many scenario factors (architecture, use case characteristics). PolyTOPS can help to design generic configurations that can work better than the state-of-the-art.

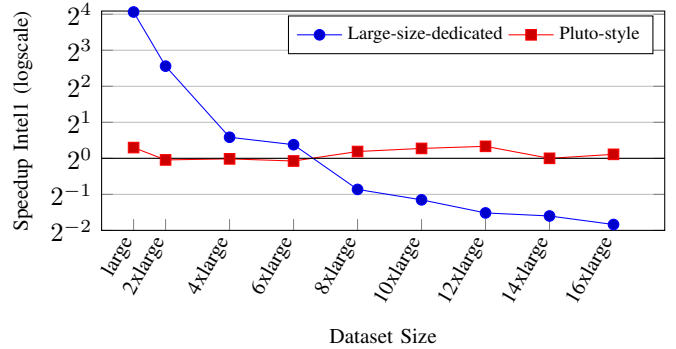
\begin{figure}[t]
  \centering{}%
    \begin{tikzpicture}%
      \begin{axis}[%
        width=0.5\textwidth,%
        height=4.5cm, 
        ymin = 0.25,%
        ymax = 17,%
        ylabel=Speedup Intel1 (logscale),%
        ylabel style = { font=\footnotesize },
        xlabel style = { font=\footnotesize },
        ytick distance = 2,%
        xtick=data,%
        xtick pos=bottom,%
        xticklabel style = { rotate = 60, anchor = east, font=\footnotesize },%
        xticklabels = {large, 2xlarge, 4xlarge, 6xlarge, 8xlarge, 10xlarge, 12xlarge, 14xlarge,         16xlarge},%
        xmin=0,%
        legend cell align={left},%
        legend style={font=\scriptsize},%
        legend columns=-1,
        xlabel=Dataset Size,%
        ymode=log,%
        log basis y={2},%
        ymajorgrids,%
      ]%

        \addplot 
	      coordinates {(1,16.69) (2,5.89) (4,1.50) (6, 1.30) (8,0.55) (10, 0.45) (12, 0.35) (14, 0.33) (16,0.28)};

        \addplot 
	      coordinates {(1,1.23) (2,0.97) (4,0.99) (6, 0.95) (8,1.14) (10, 1.21) (12, 1.26) (14, 1.00 ) (16,1.08)};
       
        \draw [black] (axis cs:-1,1) -- (axis cs:20,1);

        \addlegendentry{Large-size-dedicated};%
        \addlegendentry{Pluto-style};%
      \end{axis}%
    \end{tikzpicture}%
    \caption{Speedups of PolyTOPS compared to Pluto for Jacobi-1d using two different configurations and multiple data set sizes. The blue one is (best) dedicated configuration considered for large size. The red one is the configuration Pluto-style.}
\label{fig:JacobiSizes}
\end{figure}

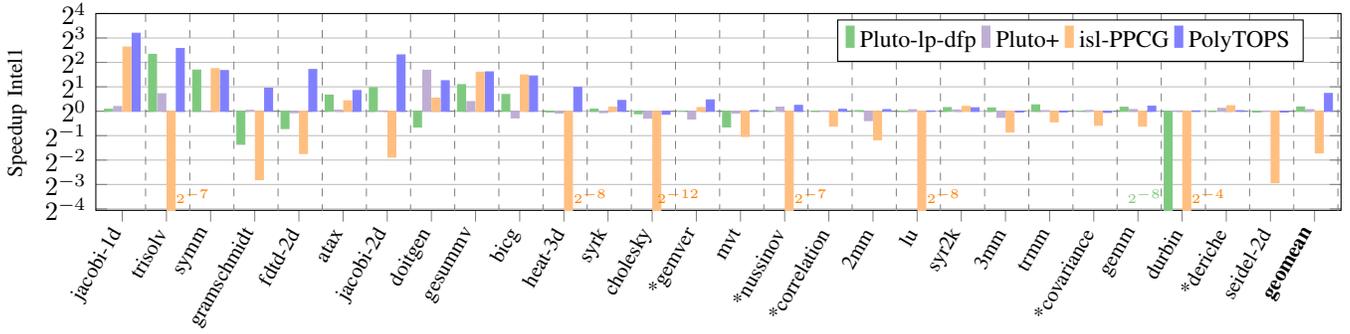
\begin{figure*}[t]
    \begin{tikzpicture}%
      \begin{axis}[%
        polybench speedup style,%
        height=4.2cm, 
        ymin = 0.06,%
        ymax = 16,%
        ytick distance = 2,%
        ylabel=Speedup Intel1,%
        ylabel style = { font=\footnotesize },
        bar width=3pt,
        legend columns=-1,
        xticklabels from table={Results/Result_Revisioned_Paper/plutolp_plutoplus.csv}{name},%
        xtick=data,%
        xticklabel style = { rotate = 60, anchor = east, font=\footnotesize },%
        ymode=log,%
        log basis y={2},%
      ]%

        \draw [help lines, dashed,ystep=100, xstep=1,  xshift=3.1mm] (0,0.006) grid(50,18);
        \addplot table [x expr=\coordindex, y=best_pluto_lp, col sep=comma]{Results/Result_Revisioned_Paper/plutolp_plutoplus.csv};%
        \addplot table [x expr=\coordindex, y=pluto_plus, col sep=comma]{Results/Result_Revisioned_Paper/plutolp_plutoplus.csv};%
        \addplot[color=orange!50] table [nodes near coords, x expr=\coordindex, y=isl, col sep=comma]{Results/Result_Revisioned_Paper/plutolp_plutoplus.csv};%
        \addplot[color=blue!50] table [x expr=\coordindex, y=polytops, col sep=comma]{Results/Result_Revisioned_Paper/plutolp_plutoplus.csv};%

        \node[font=\tiny, color=orange] at (1.6, 0.09) {$2^{-7}$};
        \node[font=\tiny, color=orange] at (10.6, 0.09) {$2^{-8}$};
        \node[font=\tiny, color=orange] at (12.6, 0.09) {$2^{-12}$};
        \node[font=\tiny, color=orange] at (15.6, 0.09) {$2^{-7}$};
        \node[font=\tiny, color=orange] at (18.6, 0.09) {$2^{-8}$};
        \node[font=\tiny, color=accent8c1] at (23.15, 0.09) {$2^{-8}$};
        \node[font=\tiny, color=orange] at (24.6, 0.09) {$2^{-4}$};

        \addlegendentry{Pluto-lp-dfp};%
        \addlegendentry{Pluto+};%
        \addlegendentry{isl-PPCG};%
        \addlegendentry{PolyTOPS};%
      \end{axis}%
    \end{tikzpicture}%
\vspace{-3mm}
\caption{Speedups (log-scale) of PolyTOPS using the kernel-specific configuration, Pluto-lp-dfp (best fusion heuristic) and Pluto+ compared to Pluto (last dev version). For the cases marked with the symbol \textbf{*}, no solution was found by any of the fusion heuristics of Pluto-lp presented in \cite{LoopFusionArticle}.}
\label{fig:ComparisonPlutoLp}
\end{figure*}

\vspace{2mm}
\subsubsection{Dataset Size Analysis}
In the context of kernel size and scheduling choices, \textit{Jacobi-1d} (similarly to \textit{trisolv}) is an example that highlights the impact of kernel size on performance. The charts in \cref{fig:BestPolybench} show that our solution outperforms Pluto for all machines. Upon closer inspection, our solution generates a simple and fully sequential code, while Pluto generates a more complex code with several conditions and complex data accesses, enabling parallelism for the inner loop through initial skewing.

The graph in \cref{fig:JacobiSizes} demonstrates how the speedup of large-size dedicated configuration (in blue) changes with different Polybench dataset sizes. We also present the results of our pluto-style configuration (in red), which is more consistent when the size changes. It is noticeable that the parallelism achieved by Pluto (and our pluto-style version) has a greater impact when the size increases. This indicates that the size of the kernel is an important factor that the scheduler should take into account, which points to a possible direction for future research. Furthermore, we want to highlight the flexibility of PolyTOPS, which can be reconfigured for all sizes, ensuring it is at least as good as Pluto, and demonstrating the power of our reconfigurability.

\subsection{Comparing scheduling tools on Polybench}
We compare PolyTOPS with Pluto+~\cite{bondhugula2016pluto-plus}, Pluto-lp-dfp~\cite{Pluto-lp} and isl-PPCG~\cite{PPCG}.
The first two works extend Pluto in several ways, while isl-PPCG is a specific version of isl-scheduler used in PPCG project.
The differences relevant to the scope of our experiments are that Pluto+ allows negative coefficients and that several fusion heuristics are available in Pluto-lp-dfp. isl-PPCG instead uses a combination of Pluto and Feautrier scheduling algorithms, and it also uses different fusion heuristics.
Speedups over Pluto are reported in \cref{fig:ComparisonPlutoLp}.
For Pluto-lp-dfp, only the highest speedup obtained from the three fusion heuristics for each code is shown~\cite{LoopFusionArticle} (except for some cases where some of the fusion heuristics did not produce a result).
The speedup shown for PolyTOPS corresponds to kernel-specific configurations. As for PolyBench, CLooG is the code generator used for all schedulers. 

In the case of \textit{doitgen}, Pluto+ outperforms PolyTOPS by enabling parametric shifting. This is a transformation by default in Pluto+, with no option to disable it.
The same solution can be obtained with PolyTOPS by enabling parametric shifting.
However, we consider it unfair to compare with Pluto, which does not allow it.

Negative coefficient support is required to find transformations for \textit{nussinov} and \textit{deriche}.
It is currently only supported in Pluto+.
Pluto, Pluto-lp-dfp and PolyTOPS only post-process the initial schedule, whereas for \textit{deriche} Pluto+ can compute a slightly better schedule.

Pluto-lp-dfp achieves a slight speedup over PolyTOPS for \textit{trmm} and \textit{3mm} due to different intra-tile optimization (post-processing).

In all other cases, PolyTOPS performs better (or similarly) than all the other versions for two reasons.
Firstly, allowing negative coefficients in Pluto+ is not beneficial for most of the Polybench cases.
Secondly, fusion heuristics Pluto-lp-dfp focus on generic high-level fusion heuristics that cannot compete (except for some cases like  \textit{trisolv} and \textit{symm}) with the kernel-specific fusion decisions from our configurations.

Regarding isl-PPCG results, cases like \textit{trisolv, gramschmidt, jacobi-2d, heat-3d, cholesky, nussinov, lu,} and \textit{seidel-2d} show big slowdowns, mainly because of fusion choices, complex skewing (caused by Feautrier cost function) that generate a complex final code. A part from that, we can see that in some other cases like \textit{jacobi-1d, symm, gesummv, and bicg} isl is capable of finding transformations that are really close (or equal) to the ones found with the best configuration of PolyTOPS.

\begin{table*}[t]

  \centering{}%
  \vspace{1mm}
  \caption{PolyMage benchmark: Timing (milliseconds) and relative speedups among PolyTOPS and the state-of-the-art schedulers (isl-PPCG, Pluto, Pluto-lp-dfp, Pluto+). For some cases and schedulers, the results are unavailable (n.a.) because of technical limitations.}
  \begin{tabular}{lcccccccccr}%
    \toprule{}%
      Benchmark &
      \begin{tabular}{@{}c@{}}PolyTOPS \\ (ms) \end{tabular} &
      \begin{tabular}{@{}c@{}}isl-PPCG \\ (ms) \end{tabular} &
      \begin{tabular}{@{}c@{}}Pluto \\ (ms) \end{tabular} &
      \begin{tabular}{@{}c@{}}Pluto-lp-dfp \\ (ms) \end{tabular} &
      \begin{tabular}{@{}c@{}}Pluto+ \\ (ms) \end{tabular} &
      \begin{tabular}{@{}c@{}}\textbf{Speedup} \\ (isl-PPCG) \end{tabular} &
      \begin{tabular}{@{}c@{}}\textbf{Speedup} \\ (Pluto-dev) \end{tabular} &
      \begin{tabular}{@{}c@{}}\textbf{Speedup} \\ (Pluto-lp-dfp) \end{tabular} &
      \begin{tabular}{@{}c@{}}\textbf{Speedup} \\ (Pluto+) \end{tabular} &
      \\
    \midrule{}%

      harris & 47 & 108 & 57  & 47 & 57 & 2.28 & 1.19 & 1 & 1.19 \\
    
      unsharp-mask & 120 & 120 & 134  &  120 & 132 & 1 & 1.10 & 1 & 1.10 \\

      camera-pipe & 88 & 177  &  n.a &  n.a. & n.a & 2.01 & n.a & n.a & n.a \\

      interpolate & 89 & 71  &  n.a &  n.a & n.a & 0.79 & n.a & n.a & n.a \\

      pyramid-blending & 74 & n.a  &  n.a &  n.a & n.a & n.a & n.a & n.a & n.a \\
    \bottomrule{}%
  \end{tabular}%
  \vspace{-3mm}
  
  \label{PolymageTable}
\end{table*}

\subsection{Comparing scheduling tools on PolyMage}
We finally compare PolyTOPS performances on the PolyMAGE \cite{polyma} benchmark suite on \textit{Intel1}. It contains 7 use cases coming from image processing. Loop-based computations and stencils characterize these codes, making them interesting scenarios for polyhedral optimizations. For our experiments, we started from the naive version of the codes provided in the benchmark and adapted them to our pipeline with several pre-processing steps. Clan~\cite{Clan} was used to transform the C++ codes into OpenScop format, and it has been adapted to support the division operation in the array indices. 

From the results in \cref{PolymageTable}, notice that many results are not available: For \textit{camera-pipe, interpolate and pyramid-blending} Pluto (in all the different versions) does not support local variables in the polyhedral representation. These are necessary to represent \texttt{if} statements using modulo and division operations, and they are also necessary for some complex accesses. isl-PPCG can handle all the cases except \textit{pyramid-blending}, where the transformation generated is empty due to some internal error.

The available results show that PolyTOPS outperforms or is on par with state-of-the-art schedulers. The codes \textit{camera-pipe}, \textit{interpolate}, and \textit{pyramid-blending} contain many statements while having a low loop dimensionality. Thus the major difficulty when optimizing them is selecting a good fusion heuristic that may enable better parallelism while remaining cache-friendly. 
For \textit{interpolate}, the performance obtained is lower than isl-PPCG because PPCG uses a more precise code generation. In our case, Cloog~\cite{Cloog} often does not take into account all directives specifying parallel dimensions for code generation, losing several parallelization opportunities. For \textit{pyramid-blending}, no code is generated by isl.

\section{Conclusion}\label{sec:conclusion}
PolyTOPS is a novel polyhedral scheduler tool that improves upon the state-of-the-art black-box polyhedral schedulers by offering an easy way to configure and tune polyhedral scheduling. 
It can adapt to various application scenarios where polyhedral optimization was previously dismissed due to the poor results of black-box schedulers.
Inputs and outputs can be expressed either as isl objects or in OpenScop format.
Thus, the output of PolyTOPS can be fed into code generation tools such as isl~\cite{isl} or CLooG~\cite{Cloog}.
PolyTOPS has been integrated into MindSpore AKG compiler~\cite{AKG}. 
The performance of PolyTOPS has been evaluated on one application scenario and on two benchmark suites. On the application scenario of \textit{\customop}~\cite{mscustomop} for an Ascend NPU, a feature of MindSpore~\cite{chen2021deep}, the configurability and flexibility of PolyTOPS led to better performance than with the isl scheduler (up to x34 speedup).
On Polybench~\cite{pouchet2012polybench} benchmark suite, we showed that simple configurations can mimic the behaviours of state-of-the-art scheduling strategies (isl, Tensor Scheduler, Pluto) and that completely new general configurations can be created with little effort, outperforming Pluto scheduler~\cite{Pluto} (x1.8 geomean speedup) using kernel-specific configurations in different CPUs.
On the Polymage benchmark suite, we have shown that PolyTOPS outperforms or is on par with other schedulers. 

This work paves the way for further research.
The design of fusion heuristics is crucial for high performance and could be an extension for PolyTOPS configurations.
Extending the currently proposed rules for fusion and defining pattern-guided fusion heuristics would be a way to enrich the existing scheduling heuristics.
Finally, more software and hardware-specific configuration extensions could prove useful: Internal heuristics for fusion, tiling adapted to the input hardware configuration and scheduling decisions based on the kernel size.


\begin{appendices}
\section{Artifact Appendix}\label{sec:artifact}

\subsection{Abstract}\label{sec:art:abstract}

This artifact provides a docker image that contains programs and scripts to generate results for
\cref{fig:BestPolybench}, \cref{fig:JacobiSizes}, \cref{fig:ComparisonPlutoLp} and \cref{PolymageTable}.
Results may differ depending on the target architecture or system.

The image contains PolyTOPS, Pluto, Pluto+, Pluto-lp, and PPCG.
Additional software such as clan, Candl, Cloog, isl and FPL are also available. 

In this artifact, you will be able to replicate the results shown in the paper and test PolyTOPS and its functionalities.

\subsection{Artifact check-list (meta-information)}\label{sec:art:checklist}

\begin{itemize}
    \item \textbf{Goal}: Reproduce results for \cref{fig:BestPolybench}, \cref{fig:JacobiSizes}, \cref{fig:ComparisonPlutoLp} and \cref{PolymageTable}
    \item \textbf{Compilation}: private
    \item \textbf{Hardware}: see \cref{sec:results:configuration}.
    \item \textbf{Metrics}: Time (Average time over the number of repetitions) in ms for PolyMage and cycles for PolyBench tests.
    \item \textbf{Output}: CSV(comma separated values) files.
    \item \textbf{How much disk space is required (approximately)?}: 6GB. 
    \item \textbf{How much time is needed to prepare workflow (approximately)?}: 1 min.
    \item \textbf{How much time is needed to complete experiments (approximately)?}: more than 12 hours for \textit{Intel1} (full experiments are required, but the timing can be tremendously reduced if excluding \textit{isl-PPCG} results as described in~\cref{sec:art:SOTA}), while around 5 hours for \textit{AMD} and \textit{Intel2} (only \cref{sec:art:polytops_experiments} needs to be tested for these 2 machines.) 
    \item \textbf{Archived?}: the artifact can be found in Zenodo~\url{ https://doi.org/10.5281/zenodo.10203989}.
\end{itemize}

\subsection{Description}\label{sec:art:description}

\subsubsection{Delivery}\label{sec:art:description:delivery}
a docker image can be found on Zenodo~\cite{PolytopsArtifact} (\url{ https://doi.org/10.5281/zenodo.10203989})

\subsubsection{Hardware dependencies}\label{sec:art:description:hardware}
see \cref{sec:results:configuration}, respectively \textit{Intel1, Intel2, AMD}.

\subsubsection{Software dependencies}\label{sec:art:description:software}
Docker v24

\subsubsection{Data Sets}\label{sec:art:description:sets}
Polybench, PolyMage


\subsection{Installation}\label{sec:art:installation}

The \texttt{\small{}docker} is published on Zenodo~\cite{PolytopsArtifact}
and can be loaded from file \texttt{\small{}polytops.tar} as follows:

\begin{lstlisting}[language=bash]
$ docker load -i polytops.tar
\end{lstlisting}
\noindent{}Upon success, image \texttt{\small{}polytops:cgo-2024} will be available.

\subsection{Experiment workflow}\label{sec:art:workflow}
A new \texttt{\small{}polytops:cgo-2024} container can be run using the following command:
\begin{lstlisting}[language=bash]
$ docker run -it --cap-add=SYS_NICE polytops:cgo-2024 
\end{lstlisting}
The image is set up so the default command is \texttt{\small{}`/bin/bash --login`}.\\
Note that the internal configuration (see \texttt{\small{}`/etc/profile.d`}) requires a login shell
for additional software to be found and executed.

Once inside the container, you can run the following commands:
\begin{lstlisting}[language=bash]
$ cd $HOME/test 
$ bash ./run_complete_artifact.sh
\end{lstlisting}

To replicate our results, we \textbf{strongly suggest} to the users to wrap test executions in the following command:
\begin{lstlisting}[language=bash]
$ sudo --login nice -n -20 bash -c "{ cd $(pwd); <test>; }"  
\end{lstlisting}
Password is \texttt{polytops}.
This command allows us to prioritize the execution of our experiments.
For instance, the previous command would become:
\begin{lstlisting}[language=bash]
$ sudo --login nice -n -20 bash -c "{ cd $(pwd);
     bash ./run_complete_artifact.sh;}"  
\end{lstlisting}
For readability, we will not rewrite it for all the following commands.

The script ''run\_complete\_artifact.sh`` will run all the scripts and generate the output timings, representing the results shown in:
\begin{itemize}
    \item \textit{PolyTOPS-results} (\cref{fig:BestPolybench}): 
    
    \texttt{\small{\$HOME/test/test\_fig2\_and\_4/fig\_2.csv}}
    \item \textit{Data-Size} (\cref{fig:JacobiSizes}): 
    
    \texttt{\small{\$HOME/test/test\_fig3/fig\_3.csv}}
    \item SOTA (\cref{fig:ComparisonPlutoLp}): 
    
    \texttt{\small{\$HOME/test/test\_fig2\_and\_4/fig\_4.csv}}
    \item PolyMage (\cref{PolymageTable}): 
    
    \texttt{\small{\$HOME/test/test\_polymage/times\_polymage.csv}}
\end{itemize}

\textbf{Notice} that the complete script is configured for \textit{Intel1}, while for \textit{Intel2} and \textit{AMD} you can refer to \cref{sec:art:polytops_experiments} that explains how to run only the specific tests in~\cref{fig:BestPolybench}.

The results can also be computed singularly for each test case. 

\vspace{1mm}

\subsubsection{PolyTOPS-results}
\label{sec:art:polytops_experiments}
To obtain the results described in \cref{fig:BestPolybench}, users can run these commands:
\begin{lstlisting}[language=bash]
$ cd $HOME/test/test_fig2_and_4/
$ bash test_fig2.sh -c $HOME/test/paper_best_configs_INTEL1 
      -n 10
\end{lstlisting}
where the option \textit{-c} specifies the root path of the PolyTOPS configuration files that we used for PolyBench cases, and \textit{-n} specifies the number of executions we want to run for each final transformation.
For the \textit{-c} option, you can select any of the following paths, depending on which experiment of \cref{fig:BestPolybench} you want to reproduce: 
\begin{itemize}
    \item For \cref{fig:BestPolybench} \textit{Intel1} machine you can use:
    
    \texttt{\small{\$HOME/test/paper\_best\_configs\_INTEL1/}}

    \item For \cref{fig:BestPolybench} \textit{Intel2} machine you can use:
    
    \texttt{\small{\$HOME/test/paper\_best\_configs\_INTEL2/}}

    \item For \cref{fig:BestPolybench} \textit{AMD} machine you can use: 
    
    \texttt{\small{\$HOME/test/paper\_best\_configs\_AMD/}}
\end{itemize}

\vspace{1mm}

\subsubsection{Data-Size}
To replicate these experiments, it is just necessary to run the following commands:
\begin{lstlisting}[language=bash]
$ cd $HOME/test/test_fig3/
$ bash test_fig3.sh -n 10
\end{lstlisting}
where \textit{-n} is an option specifying the number of executions for each program version.
The output is generated automatically in \texttt{\small{\$HOME/test/test\_fig3/fig\_3.csv}}.

\vspace{1mm}

\subsubsection{SOTA}
\label{sec:art:SOTA}
To replicate \cref{fig:ComparisonPlutoLp}, you can run the following commands:
\begin{lstlisting}[language=bash]
$ cd $HOME/test/test_fig2_and_4/
$ bash test_fig4.sh -n 10
\end{lstlisting}
where \textit{-n} specifies the number of executions for each program version.
\textbf{Notice} that in this part of the experiments, a big portion of time is taken by isl-PPCG results (see the tremendous slowdowns in~\cref{fig:ComparisonPlutoLp}). If the user wants to exclude isl from the experiment, the \texttt{\small{\$HOME/test/test\_fig2\_and\_4/test\_fig4.sh}} script can be edited at line 32, removing the \textit{isl} keyword.

\vspace{1mm}

\subsubsection{PolyMage}
To replicate the PolyMage experiments in \cref{PolymageTable}, you can run the following command:

\begin{lstlisting}[language=bash]
$ cd $HOME/test/test_polymage
$ bash test_polymage.sh
\end{lstlisting}
The \textit{output} will be available in the file\\
\texttt{\small{\$HOME/test/test\_polymage/times\_polymage.csv}}.

\subsection{Evaluation and expected result}\label{sec:art:result}
The results produced (\textit{PolyTOPS-results}, \textit{Data-Size}, \textit{SOTA}, and \textit{PolyMage}) are CSV files containing the average execution time of the different versions of the test cases and the standard deviation of the timing.

The results in the paper are equivalent for \textit{PolyMage} (\cref{PolymageTable}), while for the other charts, we calculated a speedup (compared to Pluto results) in log scale (base 2) using the following formula:

\begin{equation*}
  speedup=pluto\_time / variant\_time
\end{equation*}
where \textit{variant\_time} represents any of the variants of PolyTOPS or any other scheduler in the charts in \cref{fig:BestPolybench}, \cref{fig:ComparisonPlutoLp}, and \cref{fig:JacobiSizes}.

\subsection{Experiment customization}\label{sec:art:custom}
If you want to use our tool for custom cases, you can use the polytops command directly.
PolyTOPS supports OpenScop as input (produced by Clan) and the final code generation is done by Cloog. 
Given an input C file \textit{input.c} (that must contain the proper PRAGMA), a simple pipeline to generate an optimized version \textit{out.c} is:

\begin{lstlisting}[language=bash]
$ clan input.c | polytops --input-format=openscop 
   --tiling=true --compute-dependencies=true 
   --output-format=openscop | cloog stdin -openscop -o
   ./out.c
\end{lstlisting}
The option \textit{--help} provides a list of all the available options.

Moreover, we also provide another script in\\
\texttt{\small{\$HOME/test/scripts/single\_case.sh}}\\
that can be used to run a similar pipeline but with some extra PolyTOPS options. This script contains several extra functionalities that can be displayed with the \textit{--help} option.

\end{appendices}


\balance

\vspace{3mm}
\printbibliography{}%

\end{document}